\documentstyle[psfig,11pt]{article}

\newcommand{\bra}[1]{\mbox{$\langle #1 |$}}
\newcommand{\ket}[1]{\mbox{$| #1 \rangle$}}
\newcommand{\bracket}[2]{\mbox{$\langle {{#1}} \mathrel{ | {\vphantom
        {{#1} {#2}}} \kern-\nulldelimiterspace} {{#2}} \rangle$}}

\newcommand{\matdd}[4]{\mbox{$\left( \begin{array}{cc} #1 & #2 \\
     #3 & #4 \end{array} \right) $}}
\newcommand{\rem}[1]{}

\newcommand{\verylongrightarrow}{-\!\!\!-\!\!\!-\!\!\!\!\longrightarrow}
\newcommand{\CNOT}{{\mbox{\small\sf CNOT} }}
\newcommand{\NOT}{{\mbox{\small\sf NOT} }}
\newcommand{\DFT}{{\mbox{\small\sf DFT} }}

\newcommand{\sCNOT}{{\mbox{$\mbox{\scriptsize \sf CNOT}$} }}
\newcommand{\sNOT}{{\mbox{$\mbox{\scriptsize \sf NOT}$} }}
\newcommand{\sDFT}{{\mbox{$\mbox{\scriptsize \sf DFT}$} }}
\newcommand{\sA}{{\mbox{$\mbox{\scriptsize \sf A}$} }}
\newcommand{\ssA}{{\mbox{$\mbox{\tiny \sf A}$} }}
\newcommand{\sB}{{\mbox{$\mbox{\scriptsize \sf B}$} }}

\def\R{\hbox{$\mit I$\kern-.277em$\mit R$}}
\def\N{\hbox{$\mit I$\kern-.277em$\mit N$}}
\def\C{\hbox{$\mit I$\kern-.7em$\mit C$}}
\def\un{\leavevmode\hbox{\normalsize1\kern-4.6pt\large1}}

\title{Quantum Physics and Computers}
\author{Adriano Barenco\\
\small
Clarendon Laboratory, University of Oxford\\
\small
Parks Road, Oxford OX1 3PU, United Kingdom}
\date{}

\begin{document}

\maketitle

\baselineskip=5.5mm

\begin{abstract}

  Recent theoretical results confirm that quantum theory provides the
  possibility of new ways of performing efficient calculations.  The
  most striking example is the factoring problem.  It has recently
  been shown that computers that exploit quantum features could factor
  large composite integers. This task is believed to be out of reach
  of classical computers as soon as the number of digits in the number
  to factor exceeds a certain limit.  The additional power of quantum
  computers comes from the possibility of employing a superposition of
  states, of following many distinct computation paths and of
  producing a final output that depends on the interference of all of
  them. This ``quantum parallelism'' outstrips by far any parallelism
  that can be thought of in classical computation and is responsible
  for the ``exponential'' speed-up of computation.

  Experimentally, however, it will be extremely difficult to ``decouple'' a
  quantum computer from its environment. Noise fluctuations due to the
  outside world, no matter how little, are sufficient to drastically
  reduce the performance of these new computing devices. To control
  the nefarious effects of this environmental noise, one needs to
  implement efficient error--correcting techniques.
\end{abstract}


\section{Computation and Physics}

We are not used to thinking of computation in physical terms. We look on 
it as made up of theoretical, mathematical operations; but under close
scrutiny, effecting a computation is essentially a physical process.
Take a simple example, say ``$2+3$''; how is this trivial computation
handled by a computer? The inputs $2$ and $3$ are two abstract
quantities, and before carrying out any computation, they are encoded
in a physical system. This can take several radically different forms
depending on the computing device: voltage potentials at the gates of
a transistor on a silicon microchip, beads on the rods of an abacus,
nerve impulses on the synapse of a neuron, etc.  The computation
itself consists of a set of instructions (referred to as an {\em
  algorithm\/}) carried out by means of a physical process.
Completion of the algorithm yields a result that can be reinterpreted
in abstract terms: we observe the physical system (for instance, by
looking at the display of a calculator) and conclude that the result
is $5$.  The crucial point here is that, although $2+3$ may be defined
abstractly, the process that enables us to conclude that $2+3$ equals $5$
is purely physical.

The theory of computation has been long considered a completely
theoretical field, detached from physics. Nevertheless, pioneers such
as Turing, Church, Post and G\"odel were able, by intuition alone, to
capture the correct physical picture, but since their work did not
refer explicitly to physics, it has been for a long time falsely
assumed that the foundations of the theory of classical computation
were self--evident and purely abstract. Only in the last two decades
were questions about the {\em physics\/} of computation asked and
consistently answered~\cite{fundamentals}.  These later developments
led to a complete and thorough understanding of the physical limits of
classical computers; but they were concerned only with the {\em
  classical\/} theory of computation, for which the computing device
is supposed to obey the laws of classical physics.  This is fine as
long as one asks questions about computers we have now: any computer
that was ever built, from the oldest abacus to the latest
supercomputer, behaves indeed in a classical fashion; but we live in a
quantum world and quantum objects tend to behave quite differently
from classical ones. So what about quantum... computers?

Despite early suggestions that ``something new'' may exist when
computers are enabled to behave in a quantum mechanical way, it was not
until the seminal work of Deutsch in 1985~\cite{D85} that the foundations
of quantum computation were laid and properly formalised. In his
article, Deutsch considers the situation where  computers behave like
quantum systems and can enter highly non--classical states. These
{\em quantum computers\/} could, for instance, exist in a
superposition of states.  Each state could follow coherently a
distinct computational path and interfer to produce a final output.
This ``quantum parallelism'', achieved in a single piece of hardware,
outstrips by far any parallelism that can be thought of in classical
computers, thus {\em potentially\/} providing quantum computers with
unprecedented power.  It took indeed another decade to gain clear
evidence of the power of quantum computers and to exhibit specific
problems that were intractable on classical computers but that could
be solved efficiently on a quantum one.  The most striking example is
the {\em factorisation\/} problem.  Shor~\cite{S94} has shown recently
that using a quantum algorithm ({\em i.e.\/} an algorithm that runs on
a quantum computer) it is possible to factor large integers efficiently.

Factorisation is believed to be intractable (or at best extrememly
difficult and time--consuming) on any classical computer, and Shor's
algorithm shows for the first time that the class of problems
accessible to quantum computers includes problems that (so far) cannot
be handled efficiently by classical devices.  In fact factorisation is
not of purely academic interest only: it is the problem which
underpins the security of many classical public key cryptosystems. For
example, RSA~\cite{RSA}, the most popular public key cryptosystem
(named after the three inventors, Rivest, Shamir, and Adleman), gets
its security from the difficulty of factoring large numbers.  Hence
for the purpose of cryptoanalysis the experimental realisation of
quantum computation is a most interesting issue.  This growing
interest in the field during these last years is backed up by the
enormous experimental progress made in testing fundamentals of quantum
mechanics. In the last decade or two, it has become possible to
isolate and study single microscopic quantum systems, giving new
insights into the meaning of quantum mechanics, opening new horizons
of research and above all giving the possibility to test fundamental
ideas such as those involved in quantum computation.

This paper is organised as follows. Section 2 is concerned with the
limits of classical computation. In section 3, I introduce the basic
notions and tools of quantum computation necessary to understand the
factorisation algorithm presented in section 4. Section 5 deals with
possible ways of assembling a quantum computer out of basic elements,
while experimental realisations are presented in section 6, together
with some fundamental limitations imposed by the difficulty of
isolating a quantum system from its environment.  The final section
gives some conclusions and discusses future prospects in the field.

\section{A Brief Look at Classical Complexity}

Setting benchmarks is a useful process to understand in which sense
quantum algorithms outperform classical ones. There are rigorous ways
of defining what makes an algorithm efficient for solving a particular
problem~\cite{complexity}. For instance we can ask questions such as
``how does the memory or the time of a computation increase with the
size of the input of the problem ?''  We generally take the input size to
be the amount of information (measured in bits) needed to specify the
input.  For example, a number $N$ requires $\sim\log_2(N)$ bits of
storage on a computer (up to a fixed multiplicative factor
($\log_2(10)$), the size is the number of digits of $N$ in a decimal
system).

As an illustration, let us look at how the time needed to solve two
related problems varies with the size of the input. On one hand
consider the problem of factoring a number $N$ of size $L$ ({\em i.e.}
$L$ digits).  Factoring $N$ means finding its prime factors {\em
  i.e\/}.  finding the set of integer numbers $\{p_i\}$ such that any
$p_i$ in that set divides $N$ with the remainder $0$.  One way to calculate the
prime factors is to try to divide $N$ by $2,3,\ldots \sqrt N$ and to
check the remainder. This method is very time consuming. It requires
about $\sqrt N \approx 10^{L/2}$ divisions, hence the time it takes to
execute this algorithm increases exponentially with $L$. An as yet
hypothetical computer that can perform as many as $10^{10}$ divisions
per second would take about a second to factor a $20$ digit number,
about a year to factor a $34$ digit number and more than the estimated
age of the Universe ($10^{17}s$) to factor a $60$ digit long number.

\begin{figure}
\vspace{4mm}
\centerline{
\psfig{width=5.56in,file=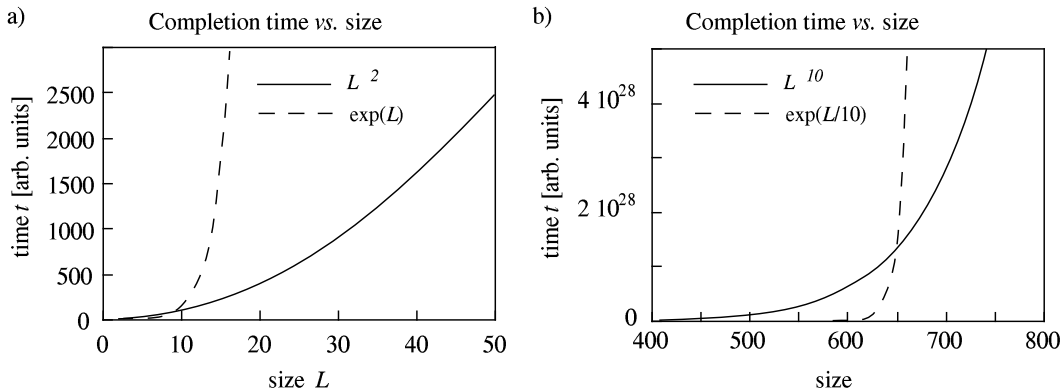}
}
\vspace{2mm}
\caption[fo0]
{\small a) Asymptotic behaviour (on the hypothetical computer) of the
  completion time (as a function of the input size) for two related
  problems. For the factorisation problem the time grows exponentially
  (dashed line), but for multiplication it only grows polynomially
  (plain line).  b) Even when the order of the exponential is small
  ($t\sim e^{\epsilon t}$, with $0<\epsilon \ll 1$, $\epsilon =0.1$ in
  the figure) and the degree of the polynomial is large ($t\sim
  t^{\eta}$ with $\eta \gg 1$, $\eta=10$ in the figure) the two curves
  will cross and above a given input size the exponential case will
  become very inefficient.  From the point of view of complexity, only
  the distinction between polynomial and exponential behaviour
  matters. }
\label{polyexp}
\end{figure}

The related problem of multiplying two $L$ digit--long numbers
together can be solved much faster. The algorithm we are taught in
school (reducing the complete multiplication in single digit
multiplications) requires $L^2$ of these basic operations. Multiplying
two $60$ digits numbers would then take only a blink of an eye on the
hypothetical computer (see Fig.~\ref{polyexp}).

For both problems, the algorithms presented are not the most
efficient: for factorisation the best algorithm is subexponential and
requires $\simeq e^{ (L^{1/3}(\log L)^{2/3} )}$
operations~\cite{lenstra}, and for multiplication
$L\log(L)\log(\log(L))$, in the limit that the numbers we multiply are
very large (more than several hundred digits)~\cite{knuth}.  Even with
these algorithms this asymmetry between the two problems remains: one
problem is solved in a time that grows (sub)exponentially with the
size of the input ({\em i.e.\/} with the number of digits of the
input), the other requires a time that grows only polynomially ({\em
  cf\/} Fig.~\ref{polyexp}b).  This asymmetry is a clear illustration
of how different problems may require a very different amount of
resources (in this case time) for obtaining the solution to a problem.

Problems for which the best algorithm runs polynomially ({\em e.g.\/}
multiplication) are said to be {\em tractable\/} and belong to what
mathematicians have called the complexity class {\sf P}, whereas, when
the time grows exponentially ({\em e.g.\/} factorisation), they are
said to be {\em intractable\/}, and belong to other classes of
complexity ({\sf NP}, {\sf EXP}, depending on other characteristics of
the problem~\cite{complexity}). The strength of this classification is
that it does not depend on the physical realisation of the computer
{\em as long as the computer obeys the laws of classical physics\/}.
Mathematicians have shown that, as long as a computer behaves
classically, it is strictly equivalent to a toy model called a {\em
  Turing Machine\/}. As practical devices, Turing Machines probably
epitomise the worst nightmare of programmers and computer scientists,
but they are an invaluable tool used by mathematicians to define and
establish complexity classes.

By showing that Turing Machines that obey the laws of quantum
mechanics could support new types of algorithms ({\em quantum
  algorithms\/}) Deutsch was able to define new complexity classes and
establish a new hierarchy~\cite{D85}.  However, it was not recognised until
recently that the class of problems that can be solved in polynomial
time with a quantum algorithm (the class {\sf QP\/}) includes problems
for which the best classical algorithm runs exponentially.
Factorisation is one of those, but before explaining how quantum
computers tackle this task, let me introduce some basic definitions.

\section{Quantum bits and pieces}

\subsection{Bits and Registers}
\label{qubits}

At the heart of a quantum computer lies the {\em quantum
  bit\/}~\cite{S95} or simply {\em qubit\/} as the natural extension
of the classical notion of bit. Instead of a simple two--state system
that can either be in state $0$ or $1$, a qubit is a quantum
two--level system, that in addition of the two eigenstate $\ket{0}$
and $\ket{1}$ (the labels are here a mere convention, but correspond
usually to the ground and excited state of the two--level system) can
be set in any superposition of the form
\begin{equation}
 \ket{\psi}=c_0\ket{0}+c_1\ket{1}.
\label{general}
\end{equation}
Any quantum two--level system is a potential candidate for a qubit,
but to help to construct a mental picture, it is a good idea to carry
a concrete, albeit somehow idealised, physical example of a qubit.  In
the following it will be useful to think of a qubit as a spin--$1/2$
particle. $\ket{0}$ and $\ket{1}$ will correspond respectively to the
spin--down and spin--up eigenstates (along a prearranged axis of
quantisation, usually set by an external constant magnetic field).

\begin{figure}
\vspace{4mm}
\centerline{
\psfig{width=5.56in,file=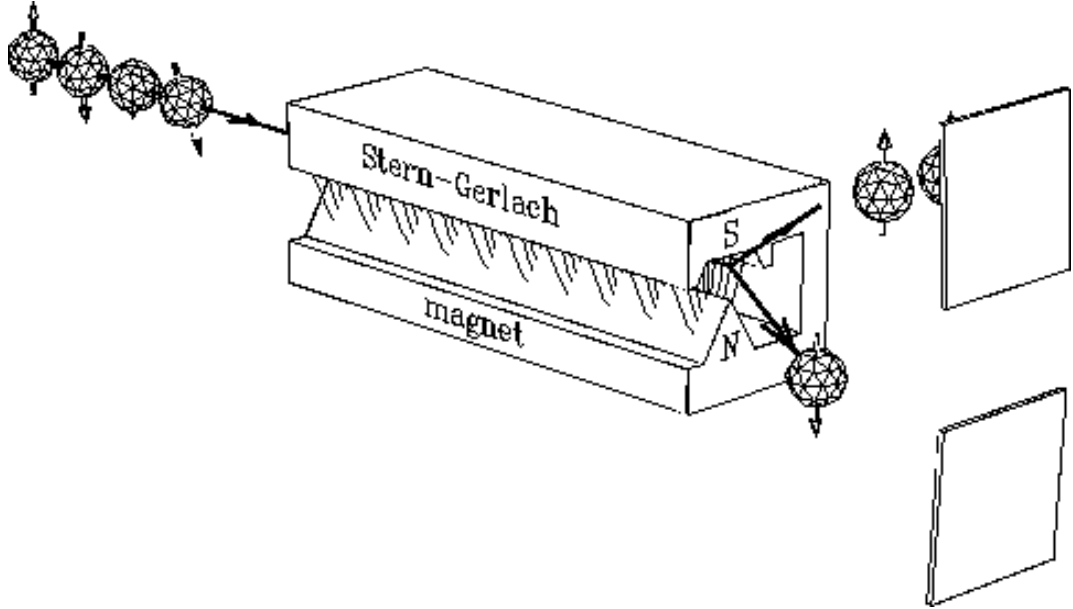}
}
\vspace{2mm}
\caption[fo1]
{\small An idealised measuring device for a quantum computer.
  Registers made out of spins, each one in a random state, pass  through a
  Stern--Gerlach magnet that performs a measurement in the spin--up
  spin--down basis ({\em i.e.\/} the basis $\ket{0}$ and $\ket{1}$).
  The spins are deflected by the device and detected by the
  detector (plates). After passing through the device, the spins point no
  longer randomly, but are in one of the two possible eigenstates.  }
\label{measurement}
\end{figure}

Although a qubit can be prepared in an infinite number of different
quantum states (by choosing different complex coefficient $c_i$'s in
Eq.~(\ref{general})) it cannot be used to transmit more than one bit
of information.  This is because no detection process can reliably
differentiate between nonorthogonal states~\cite{qdetect}.  However,
qubits (and more generally information encoded in quantum systems) can
be used in
systems developed for quantum cryptography~\cite{qcrypt},
quantum teleportation~\cite{BBCJPW93} or quantum dense
coding~\cite{BW92}.  The problem of measuring a quantum system is a
central one in quantum theory, and much attention has been and is
still devoted to this subject~\cite{peres}. In a classical computer,
it is possible in principle to inquire at any time (and without
disturbing the computer) about the state of any bit in the memory. In
a quantum computer, the situation is different.  Qubits can be in
superposed states, or can even be entangled with each other, and the
mere act of measuring the quantum computer alters its state.
Performing a measurement on a qubit in a state given by
Eq.~(\ref{general}) will return $0$ with probability $|c_0|^2$ and $1$
with probability $|c_1|^2$; but more importantly, the state of the
qubit after the measurement ({\em post--measurement state\/}) will be
$\ket{0}$ or $\ket{1}$ (depending on the outcome), and not
$c_0\ket{0}+c_1\ket{1}$.  With our spins, it is convenient to think of
the measuring apparatus as a Stern--Gerlach device into which the
qubits are sent when we want to measure them. When measuring a state
of the form of Eq.~(\ref{general}), outcomes $0$ and $1$ will be
recorded with a probability $|c_0|^2$ and $|c_1|^2$ on the respective
detector plate (see Fig.~\ref{measurement}).

We will call a collection of qubits a {\em quantum register\/}, or
simply a {\em register\/}.  As in the classical case, it can be used
to encode more complicated information.  For instance, the binary form
of $6$ is $110$ and loading a quantum register with this value is done
by preparing three qubits in state $|1\rangle \otimes |1\rangle
\otimes |0\rangle$.  In the following, we use a more compact notation:
$|a\rangle$ stands for the direct product $|a_{n-1}\rangle \otimes
|a_{n-2}\rangle \ldots|a_1\rangle \otimes|a_0\rangle$ which denotes a
quantum register prepared with the value $a=2^0 a_0+ 2^1 a_1 +\ldots
2^{n-1} a_{n-1}$. Two states $\ket{a}$ and $\ket{b}$ are orthogonal as soon as
$a\neq b$:
\begin{equation}
\bracket{a}{b}=\bracket{a_0}{b_0} \bracket{a_1}{b_1} \ldots
\bracket{a_{n-1}}{b_{n-1}},  
\end{equation}
and if $a\neq b$ at least one of the terms in the r.h.s of the above
expression is zero so that $\bracket{a}{b}=0$.

For an $n$--bit register, the most general state can be written as
\begin{equation}
  \ket{\Psi} = \sum_{x=0}^{2^n-1} c_x \ket{x}.
\label{register}
\end{equation}
Note that this state describes the situation in which several
different values of the register are present {\em simultaneously\/};
just as in the case of the qubit, there is no classical counterpart to
this situation, and there is no way to gain a complete knowledge of
the state of a register through a single measurement.

With our spin picture in mind, measuring the state of a register is
done by passing one by one the various spins that form the register in
a Stern--Gerlach apparatus and record the results
(Fig.~\ref{measurement}). For instance a two--bit register initially
prepared in the state $\ket{\psi}=1/\sqrt{2}(\ket{0}+\ket{3})$, {\em
  i.e.\/} $1/\sqrt{2}(\ket{0}\ket{0}+ \ket{1}\ket{1})$, will with
equal probability result in either two successive clicks in the
up--detector or two successive clicks in the the down--detector. The
post--measurement state will be either $\ket{0}$ or $\ket{3}$,
depending on the outcome.  A record of a click--up followed by a
click--down, or the opposite (click--down followed by click--up),
signals an experimental or a preparation error, because neither
$\ket{2}=\ket{1}\ket{0}$ nor $\ket{1}=\ket{0}\ket{1}$ appear in
$\ket{\psi}$.

\subsection{Gates}
\label{gates}

In a classical computer, the processing of information is done by
logic gates.  A logic gate maps the state of its input bits into
another state according to a {\em truth table\/}. The simplest non--trivial
classical gate is the \NOT gate, a 
one--bit gate which negates the state of the input bit: $0$ becomes
$1$ and vice--versa.  The corresponding {\em quantum gate\/} is
implemented via a unitary operation that evolves the basis states into
the corresponding states according to the same truth table. For
instance the quantum version of the classical \NOT is the unitary
operation $U_{\sNOT}$ such
\begin{equation}
\begin{array}{l}
  U_{\sNOT} \ket{0}=\ket{1} \\
  U_{\sNOT} \ket{1}=\ket{0}. 
\end{array}
\end{equation}
In quantum mechanics, the notion of gate can be extended to operations
that have no classical counterpart.  
For instance, the operation $U_{\sA}$ that evolves
\begin{equation}
\begin{array}{l}
  U_{\sA}\ket{0} = {1}/{\sqrt{2}} \;(\ket{0}+\ket{1}) \\
  U_{\sA}\ket{1} = {1}/{\sqrt{2}} \;(\ket{0}-\ket{1}),
\end{array}
\end{equation}
defines a perfectly ``legal'' quantum gate. Note that it evolves
``classical'' states into superpositions and therefore cannot be
regarded as classical.  This gate is of great utility: take an $n$--bit
quantum register initially in the state $\ket{0}$ and apply to every
single qubit of the register the gate $U_{\sA}$. The resulting state
is
\begin{equation}
\begin{array}{rl}
  \ket{\psi}&= 
U_{\sA}\otimes U_{\sA}\otimes \ldots U_{\sA}\ket{00 \ldots 0}
\\
  &= \frac{1}{\sqrt{2}}(\ket{0}+\ket{1}) \otimes
 \frac{1}{\sqrt{2}}(\ket{0}+\ket{1}) \otimes \ldots
 \frac{1}{\sqrt{2}}(\ket{0}+\ket{1}) \\ 
&=
 \frac{1}{2^{n/2}}(\ket{00\ldots0}+\ket{00\ldots1}+\ldots
 \ket{11\ldots1})
\\ 
&= \frac{1}{2^{n/2}}\sum_{x=0}^{2^n-1}\ket{x}.
\end{array}
\end{equation}
Thus, with a {\em linear\/} number of operations ({\em i.e.\/} $n$
application of $U_{\sA}$) we have generated a register state that
contains an {\em exponential\/} ($2^n$) number of distinct terms.  It
is quite remarkable that using quantum registers, $n$ elementary
operations can generate a state containing all $2^n$ possible
numerical values of the register. In contrast, in classical registers
$n$ elementary operations can only prepare one state of the register
representing one specific number.  It is this ability of creating
quantum superpositions which makes the ``quantum parallel processing"
possible.  If after preparing the register in a coherent superposition
of several numbers all subsequent computational operations are unitary
and linear ({\em i.e.\/} preserve the superpositions of states) then
with each computational step the computation is performed
simultaneously on all the numbers present in the superposition.

\subsection{Functions}
\label{functions}

Let me next describe now how quantum computers deal with functions.
Consider a function
\begin{equation}
  f:\, \{0,1,...\, 2^m-1\}\longrightarrow\{0,1,...\, 2^n-1\},
\end{equation}
where $m$ and $n$ are positive integers.  A classical device computes
$f$ by evolving each labeled input, $0,1,...\, 2^m-1$ into its
respective labeled output $f(0), f(1),...\, f(2^m-1)$. Quantum
computers, due to the unitary (and therefore reversible) nature of
their evolution, compute functions in a slightly different way.
Indeed, it is not directly possible to compute a function $f$ by a
unitary operation that evolves $\ket{x}$ into $\ket{f(x)}$: if $f$ is
not a one--to--one mapping ({\em i.e.\/} if $f(x)=f(y)$ for some $x
\neq y$), then two orthogonal kets $\ket{x}$ and $\ket{y}$ can be
evolved into the same ket $\ket{f(x)} = \ket{f(y)}$, thus violating 
unitarity.  One way to compute functions which are not one--to--one
mappings, while preserving the reversibility of computation, is by
keeping the record of the input.  To achieve this, a quantum computer
uses two registers: the first register to store the input data, the
second one for the output data.  Each possible input $x$ is
represented by $\ket{x}$, the quantum state of the first register.
Analogously, each possible output $y=f(x)$ is
represented by $\ket{y}$, the quantum state of the second register.
States corresponding to different inputs and different outputs are
orthogonal, $\langle x\vert x'\rangle =\delta_{xx'}$, $\langle y\vert
y'\rangle =\delta_{yy'}$.  The function evaluation is then determined
by a unitary evolution operator $U_f$ that acts on both registers:
\begin{equation}
  U_f\ket{x}\ket{0}=\ket{x}\ket{f(x)}.
\end{equation}

It has been  shown that as far as computational complexity is concerned,
a reversible function evaluation, {\em i.e.\/} the one that keeps
track of the input, is as good as a regular, irreversible
evaluation~[22]. This means that if a given function can be computed
in polynomial time, it can also be computed in polynomial time using a
reversible computation.

The computations we are considering here are not only reversible but
also quantum, and we can do much more than computing values of $f(x)$
one by one. We can prepare a superposition of all input values as a
single state and by running the computation $U_f$ {\em only once}, we
can compute {\em all} of the $2^m$ values $f(0), \ldots ,f(2^m-1)$,
\begin{equation}
  \ket{\psi}= U_f \left(\frac{1}{2^{m/2}}\sum_{x=0}^{2^m-1}
  \ket{x}\right)\ket{0} =
  \frac{1}{2^{m/2}}\sum_{x=0}^{2^m-1}\ket{x}\ket{f(x)}.
\end{equation}
It looks too good to be true so where is the catch? How much
information about $f$ does the state $\ket{\psi}$ contain?

As we would expect, no quantum measurement can extract all of the
$2^m$ values $f(0), f(1),\ldots ,f(2^m-1)$ {from} $\ket{\psi}$.
Imagine, for instance, performing a measurement on the first register
of $\ket{\psi}$. Quantum mechanics enables us to infer several facts:
\begin{itemize}
\item Since each value $x$ appears with the same complex amplitude in
  the first register of state $\ket{\psi}$, the outcome of the
  measurement is equiprobable and can be any value ranging from $0$ to
  $2^m-1$.
\item Assuming that the result of the measurement is $\ket{j}$, the
  {\em post--measurement\/} state of the two registers ({\em i.e.\/}
  the state of the registers after the measurement) is $\ket{\tilde
    \psi} = \ket{j} \ket{f(j)}$.  Thus a subsequent measurement on the
  second register would yield with certainty the result $f(j)$, and no
  additional information about $f$ can be gained.
\end{itemize}
However, there are more subtle measurements that provide us with
information about joint properties of all the output values $f(x)$
such as, for example, the periodicity of $f$. I will describe in the
following sections how an efficient periodicity estimate can lead to a
fast factorisation algorithm. But let me first introduce some
mathematical results in number theory that are relevant to the
problem. I shall not attempt to give a rigorous exposition nor to
provide proofs, as they can be found in most textbooks on the subject
(see for example~\cite{hardy}).

\section{Quantum Factorisation}

\subsection{Periodic functions}
\label{mathbit}

The factorisation problem can be related to finding periods of certain
functions. In particular one can show~\cite{M76} that finding factors
of $N$ is equivalent to finding the period of the function
\begin{equation}
  f_{a,N}(x) = a^x \bmod N
\end{equation}
where $a$ is {\em any} randomly chosen  number which is
coprime with $N$, {\em i.e.\/} which has no common factors with $N$
(if $a$ is not coprime with $N$, then the factors of $N$ are trivially
found by computing the greatest common divisor of $a$ and $N$).  This
function gives the remainder after the division of $a^x$ by $N$. The
function is periodic with period $r$~\cite{hardy}, which depends on
$a$ and $N$.

Knowing the period $r$ of $f_{a,N}$, we can factor $N$ provided $r$ is
even and $r \bmod N \ne -1$.  When $a$ is chosen randomly the two
conditions are satisfied with probability greater than $1/2$. The
factors of $N$ are then given by $\gcd(a^{r/2}\pm 1,N)$, the greatest
common divisor of $a^{r/2}\pm 1$ and $N$.  Fortunately, for this last
calculation, an easy and very efficient algorithm has been known since
300 BC. The algorithm, known as the Euclidean algorithm, runs in
polynomial time on a classical computer, reducing thus the problem of
factoring big numbers
to the related task of finding the period of a function.

To see how this method works, let me illustrate it with a very simple
example.  Let us try to factor $N=15$.  Firstly we select $a$, such
that $\gcd(a,N)=1$, for instance $a=7$ (with $N=15$, $a$ could be any
number {from} the set $\{2,4,7,8,11,13,14\}$).  The values of
$f_{7,15}(x)=7^x\bmod 15$ for $x=1,2,3,4,5,6,7\ldots$ are
$1,7,4,13,1,7,4\ldots$ respectively and clearly the period here is
$r=4$.  $a^{r/2}$ gives $49$ and by computing the largest common
divisors $\gcd(a^{r/2} \pm 1, N)$, we find the two factors of $15$:
$\gcd(48,15)=3$ and $\gcd(50,15)=5$.  The periods of $f_{a,15}(x)$ for
other values $a$ in the set $\{2,4,7,8,11,13,14\}$ are respectively
$\{4,2,4,4,2,4,2\}$ and in this particular example any choice of $a$
except $a=14$ leads to the correct result.  For $a=14$ we obtain
$r=2$, $a^{r/2}\equiv -1 \bmod 15$ and the method fails.

Every step of this method, except finding $r$, can be performed in
polynomial time on a classical computer.  Unfortunately, finding $r$
is actually as time consuming as finding factors of $N$ by the trial
division method since it requires us to evaluate $f_{a,N}(x)$ an
number of times exponential in $L$ on average (where $L$ is the size
of the number we want to factor, $L\simeq\log_2(N)$); however, if we
employ quantum computation, $r$ can be evaluated very efficiently.
Shor~\cite{S94} describes a quantum algorithm which provides the
period $r$ of a function and which runs efficiently ({\em i.e.\/} in
polynomial time) on a quantum computer. Let me now outline the main
features of this algorithm.

\subsection{Using quantum parallelism for factorisation.}

As was pointed out in Sect.~\ref{functions}, quantum mechanics enables
us to compute a function $f$ for different values by just applying the
corresponding unitary operator $U_f$ on a register previously set in a
superposition of orthogonal states.  Let us play this game for the
function $f_{a,N}(x)$. Since the result cannot be larger than $N$, the
output register, as defined in Sect.~\ref{functions}, should have at
least $L$ qubits. For reasons that will become clear later, we will
consider an input register of $2L$ bits. Both registers are initially
loaded with the value $0$ and the total initial state is
\begin{equation}
\ket{0}\ket{0}.
\end{equation}
We first set the input register into an equally weighted superposition
of all possible states, {from} $0$ to $2^{2L}-1$ ($\simeq N^2$), by
applying the gate $U_{\sA}$ (defined in Sect.~\ref{gates}) on each
qubit of the input register
\begin{equation}
  \frac{1}{2^L}\sum_{x=0}^{2^{2L}-1}\ket{x}\ket{0}.
\end{equation}
Applying the operator $U_{f_{a,N}}$ to this state, we obtain
\begin{equation}
  \frac{1}{2^L}\sum_{x=0}^{2^{2L}-1}\ket{x}\ket{f_{a,N}(x)}.
\end{equation}
At this stage, all the possible values of $f_{a,N}$ are encoded in the
state of the second register, but as was pointed out earlier, they are
not all accessible at the same time. On the other hand, we are not
interested in the values {\em themselves\/} but only in the
periodicity of the function.  Let me proceed now with an example to
see how this periodicity can be efficiently retrieved.

Taking the same values as in the example of the previous section ($N=15$
and $a=7$), the state of the two registers after applying
$U_{f_{7,15}}$ is
\begin{equation} 
  \frac{1}{64}\left( \ket{0}\ket{1}+ \ket{1}\ket{7}+ \ket{2}\ket{4}+
  \ket{3}\ket{13}+ \ket{4}\ket{1}+ \ket{5}\ket{7}+ \ket{6}\ket{4}+
  \ket{7}\ket{13}+\ldots \right),
\label{before}
\end{equation}
At this point, we perform a measurement on the second register.  We
take each spin that forms the second register, pass it through
our Stern--Gerlach measurement apparatus, record each outcome ($0$ or
$1$) and reconstruct the value of the register. In Eq.~(\ref{before}),
the second register encodes only the four different values $1,4,7$ or
$13$, and therefore any other measurement outcome is impossible,
unless an experimental error has occured. The state of the second
register after a measurement with outcome $j$ is
$\ket{j}$\footnote{With the Stern--Gerlach apparatus of the previous
  section, one can consider that the spins of the second register are
  ``absorbed'' by the detectors, and not available anymore after the
  measurement. For the factorisation algorithm presented here, this
  does not matter, as the second register will never be used again.
  However, knowing the outcome $j$, one could in principle easily
  reconstruct a quantum register in the state $\ket{j}$ by using, for
  instance, new spins.}.  For the first register the situation is a
bit more delicate and the post--measurement state of the first
register will be an equally weighted superposition of the states in
Eq.~(\ref{before}) for which the second register was in state
$\ket{j}$.  Table~\ref{outcomes} sums up the possible outcomes and the
post--measurement states of the two registers.
\begin{table}
\begin{equation}
\begin{array}{c|l|c}
\mbox{outcome}&\hfill \mbox{post--measurement state} \hfill &
\mbox{offset $l$} \\
\hline \hline \vspace{-2mm}
\\
1&\xi (\ket{0}+\ket{4}+\ket{8}+\;\:\ldots ) \ket{1} &0\\
4&\xi (\ket{3}+\ket{7}+\ket{11}+\ldots ) \ket{4} &3\\
7&\xi (\ket{1}+\ket{5}+\ket{9}+\;\:\ldots )\ket{7} &1\\
13&\xi (\ket{2}+\ket{6}+\ket{10}+\ldots ) \ket{13} &2\\
\end{array} \nonumber
\end{equation}
\caption[t1]{
  \small Possible outcomes of the measurement performed on the second
  register of the state of the form $\frac{1}{2^L}
  \sum_{x=0}^{2^{2L}-1} \ket{x}\ket{f_{a,N}(x)}$, with $a=4$ and
  $N=15$. The post--measurement state and the offset $l$ is also given
  for each possible outcome.}
\label{outcomes}
\end{table}
We forget now about the second register and focus only on
the state of the first one.

Let me dream for a while and imagine that quantum mechanics
enables me to {\em dictate\/} the outcome of the measurement I perform
on the second register. Imagine that I {\em decide\/} always to
obtain, say, $4$.  In this case, I would be able to prepare at will
the quantum computer in the state
\begin{equation}
\xi (\ket{3}+\ket{7}+\ket{11}+\ldots ) \ket{4}.
\end{equation}
Returning to normal rules of quantum mechanics, I could now perform a
measurement on the first register and obtain with equal probability
any of the values $3,7,11, 15\ldots$ etc. Repeating the procedure
{from} the beginning two more times, I could, with a very high
probability obtain two other distinct values, which would enable me
to find the period very easily: if in these three successive runs, I
obtain for instance $19$, $3$ and $11$, the period is easily found to
be $\gcd(19-11,11-3)=4$.

Unfortunately, dictating the result of a measurement on the second
register violates the rules of quantum mechanics.  Measurement
outcomes are probabilistic, and in my example, each allowed outcome
($1,4,7$ or $13$) is equiprobable.  In
this particular case, 
I could repeat the experiment a few times and retain only the runs for
which the outcome is $4$. However, the notion of efficiency is defined
for asymptotic behaviour and not for particular cases. The real
question is to know how this technique will perform for increasing
$N$'s.  Quite clearly, it does not look promising; in a general case,
when the period is $r$, there are $r$ possible different outcomes.
Since $r$ also grows exponentially with $L$ (the size of $N$), the
approach that consists of repeating the quantum computation over and
over again until measuring in the second register at least three times
the same value (in order then to perform a measurement on the first
register and find the factor via a greatest common divisor
calculation) is inefficient.

An additional ingredient is needed to make the quantum algorithm
polynomially efficient.  Whatever the outcome of the measurement was,
the first register is left in an equally weighted superposition of the form
\begin{equation}
  \ket{\psi}=\xi \sum_{j=0}^{\lfloor 2^{2L}/r \rfloor}  \ket{jr+l},
\label{firstreg}
\end{equation}
with $r$ being the period of $f_{a,N}(x)$,  $l$ an offset value and
$\xi$ an appropriate normalisation factor. 
({\em cf\/}.  Fig.~\ref{cxcy}a and Table~\ref{outcomes}).  Regardless
of the outcome of the measurement, the period $r$ of the function
$f_{a,N}$ is always reflected in the post--measurement state of the
first register. But it is not readily accessible, as the offset $l$
depends probabilistically on the outcome of the measurement. It is
nevertheless possible to get rid of this irritating offset by using a
quantum equivalent of the classical Fast Fourier Transform. This
operation is known as the Discrete Fourier Transform ({\small \sf DFT}).

\subsection{Discrete Fourier Transform}

Consider the unitary operation $U_{\sDFT}$ that acts on a quantum
register and effects
\begin{equation}
  U_{\sDFT}\ket{x}=\frac{1}{2^L}\sum_y^{2^{2L}-1}\exp(2\pi i
  \frac{xy}{2^{2L}}) \ket{y},
\end{equation}
where $2L$ is the size of the register.
The reason for calling this particular unitary transformation the
discrete Fourier transform becomes obvious when we notice that in the
transformation
\begin{equation}
  U_{\sDFT} \sum_{x=0}^{2^{2L}-1} c_x\ket{x} =\sum_y c_y\ket{y}
\end{equation}
the coefficients $c_y$ are the discrete Fourier transforms of $c_x$'s
{\em i.e.\/}
\begin{equation}
  c_y= \frac{1}{2^L}\sum_{x=0}^{2^{2L}-1} \exp(2\pi i \frac{xy}{2^{2L}})
  c_x.
\end{equation}

\begin{figure}
\vspace{4mm}
\centerline{
\psfig{width=5.56in,file=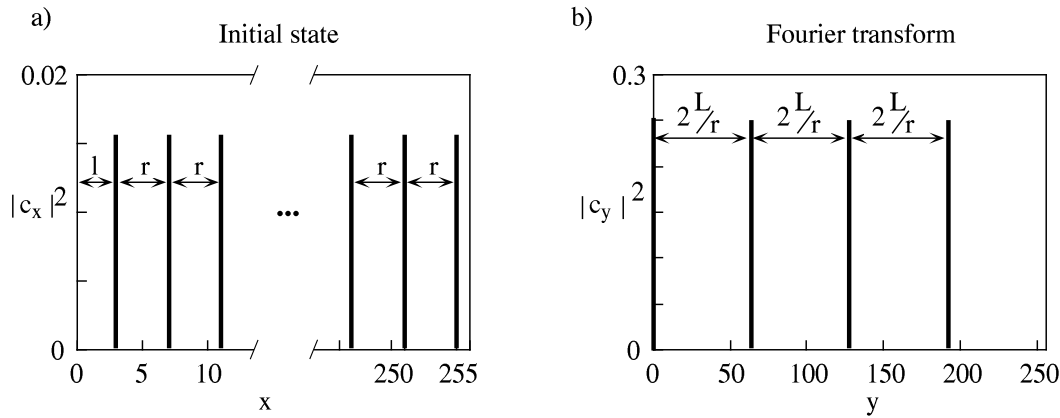}
}
\vspace{2mm}
\caption[fo2]
{\small The \DFT on an input state of the form $\sum_x c_x \ket{x}$
  results in $\sum_y c_y \ket{y}$. When the input state is periodic,
  as in (a), the effect of the \DFT is to eliminate the eventual
  offset $l$ and to invert the period $r \rightarrow 2^L/r$ (b). In
  the figure $L=8$, $l=3$ and $r=4$.  In this particular case the
  period $r$ divides exactly $2^L$, resulting thus in a ``clean''
  transformation. Appendix~\ref{appendDFT} describes a more general
  case where $r$ does not divide $2^L$, in this situation a slight
  spread occurs in the peaks of the Fourier transformed state ({\em
    cf.\/} Fig.~\ref{nonexactcxcy}).  }
\label{cxcy}
\end{figure}

The strength of the \DFT lies in the fact that when it acts on a
periodic state of the form of Eq.~(\ref{firstreg}), it will wipe out
the offset $l$, and transform it into a phase factor that does not
affect the probabilistic outcome of a later measurement on the
register.  Appendix~\ref{appendDFT} shows how the \DFT on a $2L$--bit
register maps a state of period $r$ into a state of period $2^{2L}/r$.
When $r$ divides exactly $2^{2L}$,  the resulting state has a nice
closed form ({\em cf.\/} Fig.~\ref{cxcy}):
\begin{equation}
  \ket{\phi_{out}} = \frac{1}{\sqrt{r}} \sum_{j=0}^{r-1} \exp (2\pi i
  lj/r) \; \ket{j 2^{2L}/r}.
\label{afterdft}
\end{equation}
A more careful analysis is required when $r$ does not divide $2^{2L}$ (see
Appendix~\ref{appendDFT}). Even in this more general case, the \DFT
retains the features illustrated in the particular situation above: it
``inverts'' the periodicity of the input ($r\rightarrow 2^{2L}/r$) and
it has this translation invariance property which washes out the
offset $l$ (Fig.~\ref{cxcy}b). Thus, by effecting $U_{\sDFT}$ on
states of the form of Eq.~(\ref{firstreg}) with different $l$, we
always end up with a state for which neither the outcome of a
measurement, nor its probability depend on $l$ anymore. 

In the previous section, I showed how to construct a state of a
register with a periodic superposition and an arbitrary offset.
Combining this method with a {\small \sf DFT}, it is now possible to
retrieve efficiently the ``inverted'' period $2^{2L}/r$ ({\em cf.\/}
Fig.~\ref{cxcy}), {from} it the period $r$, and finally the factors of
$N$.

\subsection{Randomised algorithms}

\begin{figure}
\vspace{4mm}
\centerline{
\psfig{width=2.7in,file=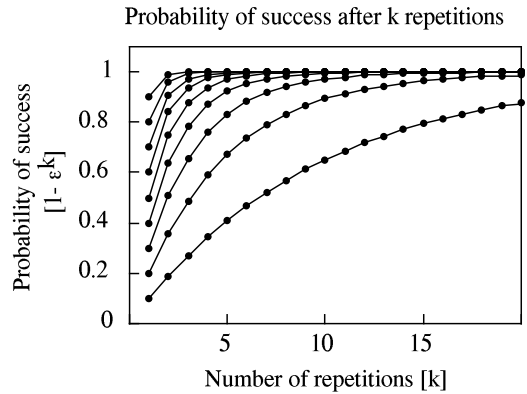}
}
\vspace{2mm}
\caption[fo3]
{\small 
Probability of having at least one successful run after $k$ runs of a
randomised algorithm with probability of success $\epsilon$. The
various curves illustrate different values of $\epsilon$, ranging from
$\epsilon=0.9$ (bottom curve), to $\epsilon=0.1$ (top curve).
}
\label{randomised}
\end{figure}

Shor's algorithm is a randomised algorithm which runs successfully
only with probability $1-\epsilon$. We know when it is successful: it
produces a candidate factor of $N$ which can be checked by a trial
division to see whether the result is indeed a factor or not. This
check can be effected in polynomial time as it just involves a
division. If $\epsilon > 0$ is independent of the input $N$, by
repeating the computation $k$ times, we get probability $1-\epsilon^k$
of having at least one success. This can be made arbitrarily close to
$1$ by choosing a fixed $k$ sufficiently large (see
Fig.~\ref{randomised}).  Furthermore, if a single computation is
efficient, then repeating it $k$ times will also be efficient since
$k$ is independent of $N$.  Thus the success probability of any
efficient randomised algorithm of this type may be amplified
arbitrarily close to $1$ while retaining its efficiency.  Indeed we
may even let the success probability $1-\epsilon$ decrease with $N$ as
$1/\mbox{poly}(\log N)$ and $k$ increase as $\mbox{poly}(\log N)$ and
still retain efficiency while amplifying the success probability as
close to $1$ as desired.  Shor's quantum factoring algorithm is of
this type; it is based on an efficient algorithm which provides a
factor of the input $N$ with a probability which decreases as
$1/\mbox{poly}(\log N)$.

The randomness in the algorithm is due to certain mathematical results
concerning the distribution of prime and coprime numbers.  For
instance, for some values of the initial number $a$, the algorithm
will fail, even if $a$ coprime with $N$ ({\em cf.\/} Sect.~\ref{mathbit}).  Also if
we abandon the assumption that $r$ divides $2^{2L}$ (very unlikely and
adopted in this section only for pedagogical purposes) the \DFT of
$c_x$ will not produce sharp maxima as in Fig.~\ref{cxcy}.  which may
contribute to possible errors while reading $y$ {from} the register.
Subsequent estimation of $r$ is calculated using additional
mathematical approximation techniques (continued fraction expansion,
see for instance~\cite{EJ96}).

If we try to factor larger and larger numbers $N$ it is enough to
repeat the computation a number of times that grows polynomially with
$L$ to amplify the probability of success as close to $1$ as we wish.
This gives an efficient determination of $r$ and an efficient method
of factoring any $N$.

\section{Towards Quantum Networks}

In the previous sections I have specified unitary operations by
describing how they affect the state of the registers on which they
act. I have not given any indications of how to implement them.  These
operations are usually quite complex.  For instance the \DFT on a
register of $L$ qubits is an operation that acts on a $2^L$
dimensional state; the mere task of writing down its matrix would take
an exponential time in $L$.  The method for implementing these unitary
operations depends on the computational paradigm that is to be used.
The only requirement being, of course, that neither the size of the
implementation nor the time necessary to perform the operation should
grow faster than a polynomial in the size of the problem.

\begin{figure}
\vspace{4mm}
\centerline{
\psfig{width=6in,file=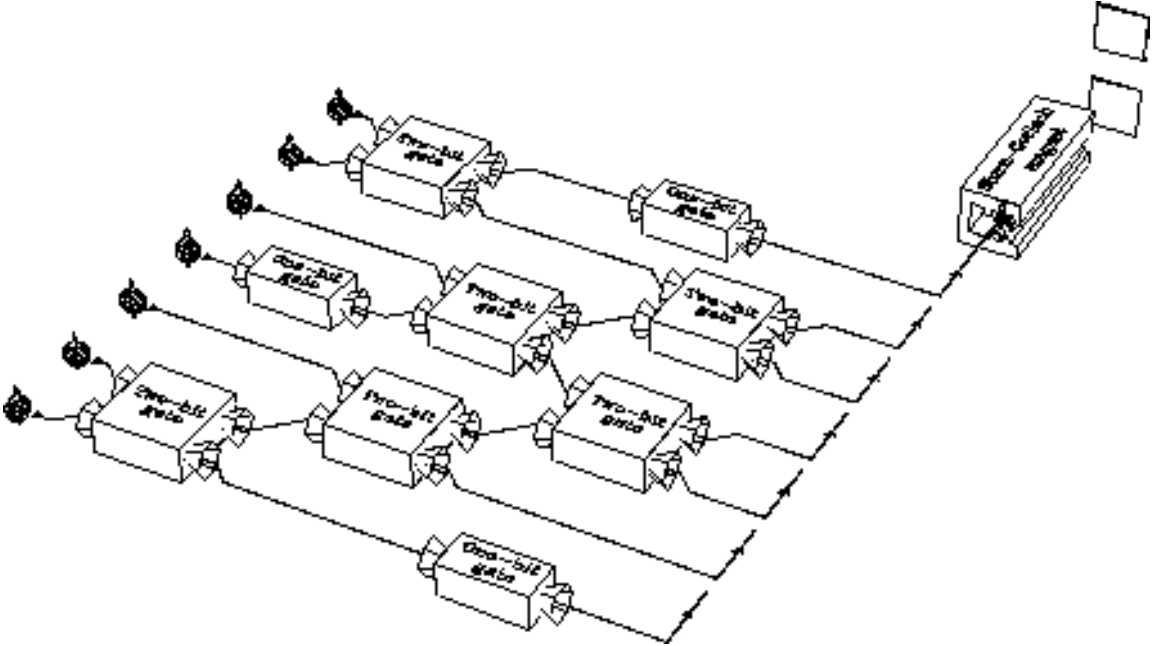}
}
\vspace{2mm}
\caption[fo4]
{\small An idealised picture of what a quantum network could look like.
  Spins enter a network of gates (on the left) and ``fly'' from one
  gate to another, where they interact with other qubits (two--bit
  gates) or are just individually manipulated (one--bit gates).
  Eventually, at the end of their journey, they are measured by a
  Stern--Gerlach apparatus.}
\label{flying}
\end{figure}

Deutsch~\cite{D89} described {\em quantum networks\/} as a possible
way to effect complex unitary operations. Quantum networks are
composed of elementary logic gates connected together by wires. The
only purpose of the wires is to transfer a quantum state from the
output of a gate to the input of another one (and eventually to a
measurement device).  Of course, just as it is the case for quantum
gates, the nature of these wires depends on the technological
realisations of the qubit.  For instance, wires could hypothetically
be the trajectories of flying spins: two spins may have their
trajectory inflected, enter a zone in which they interact together (a
two--bit gate), fly apart again, enter another zone in which they
interact with other qubits, etc. (see Fig.~\ref{flying}). The
fundamental idea underlying quantum networks is to decompose complex
unitary operations acting on several qubits into a sequence of simple
one-- and two--bit gates. Other paradigms to implement quantum
computation involve for instance quantum cellular automata, but so far
they have not proved to be tools as valuable as the idea of quantum
networks. I will not discuss them here, and the interested reader can
refer to the literature on the subject~\cite{margolus}.

Deutsch showed that there exists a universal three--bit quantum gate
{from} which any quantum computation, {\em i.e.\/} any unitary
operation on any finite number of qubits, can be built by a suitable
network consisting only of copies of this gate. This result has been
improved upon since then~\cite{twobit} and we now know that almost any
non--trivial two--bit gate is universal~\cite{universal}.  Much
attention has also been devoted to the efficient construction of more
complex quantum gates~\cite{gang9SD95}, and to specific networks, such
as the one that effects the modular exponentiation required in the
first part of the factorisation algorithm~\cite{VBE96}. In the
following, I will illustrate how the quantum discrete Fourier
transform discussed above can be implemented as a network consisting
of only one-- and two--bit gates.

\setlength{\unitlength}{0.030in} Consider again the one--bit gate
$U_{\sA}$ of Sect.~\ref{gates} performing the unitary transformation
\begin{equation}
\begin{array}{l}
 U_{\sA} \ket{0}=\frac{1}{\sqrt{2}} \;(\ket{0}+\ket{1}) \\
 U_{\sA} \ket{1}=\frac{1}{\sqrt{2}}\; (\ket{0}-\ket{1}).
\end{array}
\mbox{\hspace{3cm}} \mbox{
\begin{picture}(30,0)(0,15)
  \put(0,15){$q$} \put(5,15){\line(1,0){5}} \put(20,15){\line(1,0){5}}
  \put(10,10){\framebox(10,10){\sf A}}
\end{picture}
}
\end{equation}
The diagram on the right provides a schematic representation of the
gate acting on a qubit $q$. Consider also the two--bit $U_{\sB(\phi)}$
gate acting on qubits $q_1$ and $q_2$ and performing the operation
\begin{equation}
  \begin{array}{l}
    U_{\sB(\phi)}\ket{00}=\ket{00}  \\
    U_{\sB(\phi)}\ket{01}=\ket{01}  \\
    U_{\sB(\phi)}\ket{10}=\ket{10}  \\
    U_{\sB(\phi)}\ket{11}=e^{i\phi}\ket{11}.
   \end{array}
   \mbox{\hspace{1.5cm}} \mbox{
\begin{picture}(25,0)(0,20)
  \put(-2,15){$q_2$} \put(-2,30){$q_1$} \put(5,15){\line(1,0){5}}
  \put(20,15){\line(1,0){5}} \put(5,30){\line(1,0){20}}
  \put(15,30){\circle*{3}} \put(15,20){\line(0,1){10}}
  \put(10,10){\framebox(10,10){$\phi$}}
\end{picture}
}
\equiv
\mbox{\hspace{0.2cm}} 
\mbox{
\begin{picture}(30,0)(0,20)
  \put(-2,15){$q_2$} \put(-2,30){$q_1$}
 \put(5,30){\line(1,0){5}}
  \put(20,30){\line(1,0){5}}
 \put(5,15){\line(1,0){20}}
  \put(15,15){\circle*{3}} \put(15,15){\line(0,1){10}}
  \put(10,25){\framebox(10,10){$\phi$}}
\end{picture}
}
\end{equation}
The gate $U_{\sB(\phi)}$ performs a conditional phase shift, {\em
  i.e.\/} a multiplication by a phase factor $e^{i\phi}$ only if the
two qubits are both in their $\ket{1}$ state. The three other basis
states are unaffected.

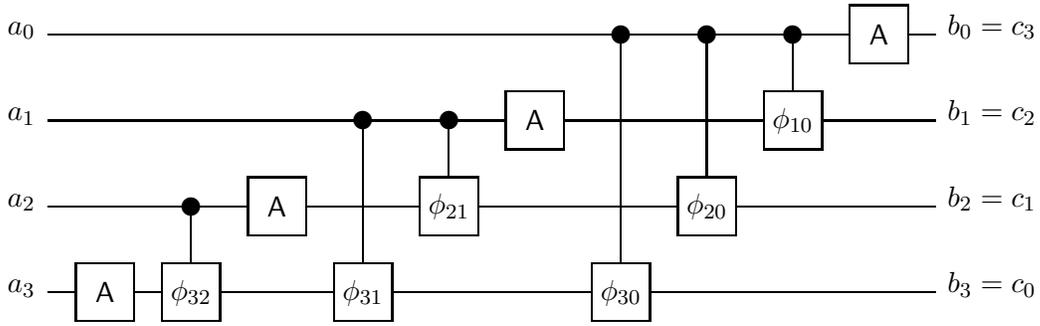
\begin{figure}
\centerline{
\begin{picture}(175,65)
  \put(-2,15){$a_3$} \put(-2,30){$a_2$} \put(-2,45){$a_1$}
  \put(-2,60){$a_0$} \put(10,10){\framebox(10,10){\sf A}}
  \put(25,10){\framebox(10,10){$\phi_{32}$}}
  \put(40,25){\framebox(10,10){\sf A}}
  \put(55,10){\framebox(10,10){$\phi_{31}$}}
  \put(70,25){\framebox(10,10){$\phi_{21}$}}
  \put(85,40){\framebox(10,10){\sf A}}
  \put(100,10){\framebox(10,10){$\phi_{30}$}}
  \put(115,25){\framebox(10,10){$\phi_{20}$}}
  \put(130,40){\framebox(10,10){$\phi_{10}$}}
  \put(145,55){\framebox(10,10){\sf A}} \put(5,15){\line(1,0){5}}
  \put(20,15){\line(1,0){5}} \put(35,15){\line(1,0){20}}
  \put(65,15){\line(1,0){35}} \put(110,15){\line(1,0){50}}
  \put(5,30){\line(1,0){35}} \put(50,30){\line(1,0){20}}
  \put(80,30){\line(1,0){35}} \put(125,30){\line(1,0){35}}
  \put(5,45){\line(1,0){80}} \put(95,45){\line(1,0){35}}
  \put(140,45){\line(1,0){20}} \put(5,60){\line(1,0){140}}
  \put(155,60){\line(1,0){5}} \put(30,30){\circle*{3}}
  \put(30,20){\line(0,1){10}} \put(60,45){\circle*{3}}
  \put(60,20){\line(0,1){25}} \put(75,45){\circle*{3}}
  \put(75,35){\line(0,1){10}} \put(105,60){\circle*{3}}
  \put(105,20){\line(0,1){40}} \put(120,60){\circle*{3}}
  \put(120,35){\line(0,1){25}} \put(135,60){\circle*{3}}
  \put(135,50){\line(0,1){10}}
\put(162,60){$b_0=c_3$}
\put(162,45){$b_1=c_2$}
\put(162,30){$b_2=c_1$}
\put(162,15){$b_3=c_0$}
\end{picture}
}
\caption[fo5]{\small
  Network effecting a \DFT on a four--bit register, the phases that
  appear in the operations $U_{\sB(\phi_{jk})}$ are related to the
  ``distance'' of the qubits upon which $U_{\sB}$ acts, namely
  $\phi_{jk}=\pi/2^{j-k}$. The network should be read from the left to
  the right: first the gate {\sf A} is effected on the qubit $a_3$, then
  $B(\phi_{32})$ on $a_2$ and $a_3$, and so on.}
\label{network}
\end{figure}

The \DFT on a register of any size can be implemented using only these
two gates.  For example, consider a four--bit register with qubits
$a_0, \ldots a_3$.  The network in Fig.~\ref{network} follows step by
step the classical algorithm of a \DFT \cite{knuth}, and performs the
operation
\begin{equation}
  \ket{a} \longmapsto \frac{1}{\sqrt{2^4}} \sum_{c=0}^{2^4-1}
  \exp(2\pi iac/2^4) \ket{b},
\end{equation}
where $\ket{b}$ represents the value $c$ read {\em reversing\/} the
order of the bits {\em i.e.\/}
\begin{equation}
  b=\sum_{i=0}^3 2^i c_{3-i} \; \; \; \mbox{with $c_k$ given by} \; \;
  \; c=\sum_{k=0}^3 2^k c_k.
\end{equation}
A trivial extension of the network following the same sequence pattern
of gates on $L$ qubits gives the general {\small \sf DFT}.  In this
case the transformation requires $L$ operations $U_{\sA}$ and
$L(L-1)/2$ operations $U_{\sB(\phi)}$, in total $L(L+1)/2$ elementary
operations. Thus the quantum \DFT can be performed efficiently.
Moreover, it can even be simplified~\cite{C94}: in a general network
for {\small \sf DFT},
the operations $U_{\sB(\phi)}$ that involve distant
qubits $a_j$ and $a_k$, {\em i.e.\/} qubits for which $|j-k|$ is large
(and therefore $\phi=\pi/2^{k-j}$ approaches zero), are close to
unity.  Therefore when performing the quantum \DFT on registers of
size $L$, one can neglect operations $U_{\sB(\phi)}$ on distant qubits (more
precisely on qubits $a_j$ and $a_k$ for which $|j-k|>\log_2(L)+2$) and
still retrieve the periodicity of coefficients $c_x$
with high probability~\cite{BETS96}.

The network of gates for the quantum \DFT enables the efficient
implementation of the second part of Shor's algorithm.  The first part
requires an efficient quantum evaluation of the function $f_{a,N}(x) =
a^x \bmod N$.  The computation of $f_{a,N}(x)$ is ``easy'' {\em
  i.e\/}. the number of elementary operations does not grow faster
than a polynomial in the size of the input.  The respective network is
constructed by combining networks which perform additions and
multiplications in a reversible and unitary way~\cite{VBE96}.

\section{Practicalities}

It remains an open question which technology will be employed to build
the first quantum computers.  The conditional quantum dynamics which I
alluded to when introducing the operation $U_{\sB}$ can be implemented
in many different ways, ranging {from} Ramsey atomic
interferometry~\cite{BN94}, interacting electrons in quantum
dots~\cite{BDEJ95} to ions in ion traps~\cite{ions} and atoms coupled
to high finesse optical resonators~\cite{caltech}.

In the following I will present a possible scheme to implement the
simple two--bit quantum gate ``controlled--{\small \sf NOT}''
({\small\sf CNOT})~\cite{BDEJ95}.  Its effect on the basis states is
\begin{equation}
  \begin{array}{l}
    U_{\sCNOT}\ket{00}=\ket{00}  \\
    U_{\sCNOT}\ket{01}=\ket{01}  \\
    U_{\sCNOT}\ket{10}=\ket{11}  \\
    U_{\sCNOT}\ket{11}=\ket{10}.
   \end{array}
\end{equation}
The gate effects a logical \NOT on the second qubit (target bit), if
and only if the first qubit (control bit) is in state $1$.  Two
interacting magnetic dipoles (for instance, the spins we have
considered in the previous sections), sufficiently close to each
other, can be used to implement this operation. Under carefully
chosen conditions~\cite{cohen}(complement $\mbox{B}_{\mbox{\scriptsize
    XI}}$) the total hamiltonian of this system can be written
\begin{equation}
 H=H_1+H_2+H_{\mbox{\scriptsize coupl}}
\label{hamilto}
\end{equation}
with
\begin{equation}
\begin{array}{l}
H_1=\hbar \omega_1 S_{1,z} \\
H_2=\hbar \omega_2 S_{2,z} \\
H_{\mbox{\scriptsize  coupl}}= 4 \hbar \Omega  S_{1,z} S_{2,z},
\end{array}
\end{equation}
where $S_{1,z}$ and $S_{2,z}$ are the $z$ components of the spin
operators for spin $1$ and $2$ respectively, so that $S_{1,z}
\ket{0,\cdot}=-1/2 \hbar \ket{0,\cdot}$ and $S_{1,z}
\ket{1,\cdot}=+1/2 \hbar \ket{1,\cdot}$ (similar relations hold for
$S_{2,z}$). $\omega_i=\gamma_i B_i$ depends on the gyromagnetic ratio
$\gamma_i$ of spin $i$ and of the magnetic field $B_i$ (along the $z$
axis) experienced by spin $i$. We will suppose that either the magnetic
fields or the gyromagnetic ratios are different for each qubit so that
$\omega_1\neq\omega_2$. Finally $\Omega$ is a coupling factor that
depends, among others, on the distance between the two spins and their
relative orientation. The form of the above hamiltonian is valid under
the assumption that the coupling $\Omega$ is small enough to be
regarded as a perturbation.

Without coupling ($\Omega=0$, for instance when both spins are
sufficiently far apart), each spin can be selectively flipped by a
resonant electromagnetic field of appropriate duration and
intensity: shining a pulse at frequency $\omega_1$ on both spins will
affect only spin $1$, switching it from state $\ket{0,\cdot}$ (spin down) to
$\ket{1,\cdot}$ (spin up), and vice versa. Similarly, a pulse of frequency
$\omega_2$ will only affect spin $2$. 

\begin{figure}
\vspace{4mm}
\centerline{
\psfig{width=4.56in,file=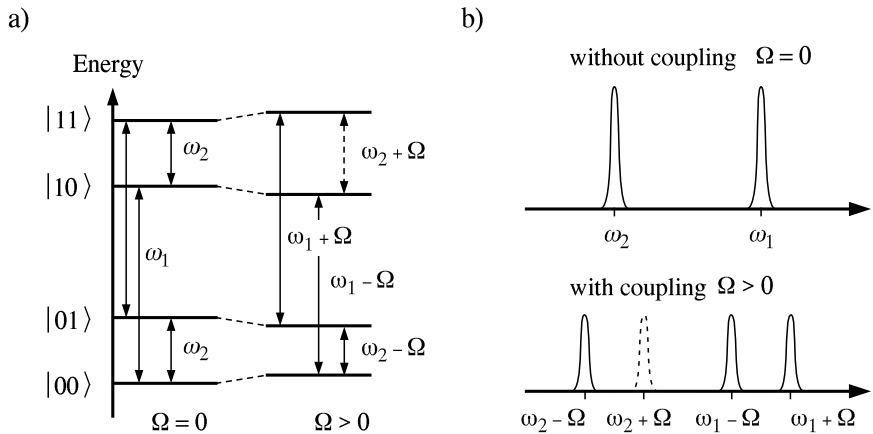}
}
\vspace{2mm}
\caption[fo6]
{\small (a) Eigenenergies of the basis state of two spins without and
  with coupling. (b) Resonant frequencies of the two spins without and
  with coupling. When there is no coupling, the two frequencies are
  made different for instance, by putting each spin in a magnetic
  field of slightly different intensity. With coupling, each resonant
  frequency is split into two frequencies; if the splitting is
  sufficient, it is possible to select specific transitions. The
  dotted transition corresponds to the \CNOT operation, in which only
  states $\ket{10}$ and $\ket{11}$ are affected.  }
\label{cnotfig}
\end{figure}

When both spins are close enough to interact ($\Omega > 0$), the
situation is more complicated and one needs to find the eigenstates of
the hamiltonian $H$. Fortunately, $H$ remains diagonal in the basis
$\{ \ket{00}, \ket{01}, \ket{10}, \ket{11} \}$ and the energies of the
eigenstates are shifted by $\pm \hbar \Omega$, depending on the state
({\em cf.\/} Fig.~\ref{cnotfig}a). By carefully selecting a resonant
frequency, it is now possible to induce selective switching of one
spin depending on the state of the other. Fig.~\ref{cnotfig}
illustrates how a pulse of frequency $\omega_2+\Omega$ induces a
switching betweeen states $\ket{10}$ and $\ket{11}$ only, leaving
states $\ket{00}$ and $\ket{01}$ unaffected, implementing thus the
\CNOT operation.

Obviously the above description is rather simplistic and does not
include any unwelcome effects; however, the form of the hamiltonian
$H$ and of the coupling $ H_{\mbox{\scriptsize coupl}}$ appears in
many radically different contexts (excitons or electrons in quantum
dots~\cite{BDEJ95}, interacting Cooper pairs in superconducting
islands~\cite{devoret}), and is general enough to make this setup
worth mentioning.

\subsection{Coupling with the environment: the decoherence problem.}

In order to perform a successful quantum computation, one has
to maintain a coherent unitary evolution until the completion of the
computation.  Technically it is not possible to ensure that a quantum
register is completely isolated from the environment. This remnant
coupling induces decay and decoherence processes, both of which
 drastically reduce the performance of a quantum computer, even when
the coupling is very weak. Decay is a process by which a quantum
system dissipates energy in the environment. For a spin, it is for
instance a transition from $\ket{1} \longrightarrow \ket{0}$,
accompanied by the emission of a photon of appropriate wavelength.
Decoherence is a subtler phenomenon that involves no exchange of
energy with the environment~\cite{zurek}. Its effect is to scramble
the relative phase of the various parts of a quantum superposition.
Decoherence occurs in most cases on a much faster timescale than
decay, and therefore, I will focus on this kind of processes.

Decoherence can be more easily understood if we formalise it in the
language of density operators, rather than in the more familiar Dirac
notation. When a quantum system is in a pure state, it can be
equivalently described by a ket $\ket{\psi}$ or by a density operator
$\rho=\ket{\psi}\bra{\psi}$.  The characteristic effect of decoherence
is to destroy the off--diagonal elements of the density operator; the
system evolves into a ``mixed state''~\cite{cohen} for which the ket
notation is no longer  suitable. To see how this can affect quantum
computers, let me first consider the very simple situation in which a
qubit initially in the state $\ket{0}$ undergoes successively and
without decoherence two operations $U_{\sA}$ (as introduced in
Sect.~\ref{gates}):
\begin{equation}
  \ket{\psi_{in}}=\ket{0} \stackrel{U_{\ssA}}{\longrightarrow}
  \frac{1}{\sqrt{2}}(\ket{0}+\ket{1})
  \stackrel{U_{\ssA}}{\longrightarrow} \ket{0} = \ket{\psi_{fin}}.
\end{equation}
In a density matrix formulation, this sequence can be written (in the
basis ${\cal B}=\{ \ket{0}, \ket{1} \}$)
\begin{equation}
  \rho_{in}=\matdd{1}{0}{0}{0} \stackrel{U_{\ssA}}{\longrightarrow}
  \frac{1}{2} \matdd{1}{1}{1}{1}
  \stackrel{U_{\ssA}}{\longrightarrow} \matdd{1}{0}{0}{0} = \rho_{fin}.
\end{equation}
A measurement of the final state would yield $0$ with probability one.
Let me suppose now that decoherence occurs in between the two
operations $U_{\sA}$ and wipes out completely the off--diagonal
elements (this is of course an oversimplification, and one should
rather picture decoherence as a continuous  process
that progressively eliminates the off--diagonal elements).  In this case,
the sequence of operations reads
\begin{equation}
  \rho_{in}=\matdd{1}{0}{0}{0} \stackrel{U_{\ssA}}{\longrightarrow}
  \frac{1}{2} \matdd{1}{1}{1}{1} \stackrel{\mbox{ \tiny \sf
      DECO.}} {\verylongrightarrow} \frac{1}{2}
  \matdd{1}{0}{0}{1} \stackrel{U_{\ssA}}{\longrightarrow} \frac{1}{2}
  \matdd{1}{0}{0}{1} = \tilde\rho_{fin},
\end{equation}
and $\tilde\rho_{fin}$ no longer  represents  a qubit in the state
$\ket{0}$, but rather a {\em statistical mixture\/} of the states
$\ket{0}$ and $\ket{1}$. Performing a measurement on the qubit would
now return either $0$ or $1$ with equal probability; thus decoherence
affects the probability distribution of the possible outcomes of a
computation.

The onset of decoherence is actually more complex, and to a large
extent depends on the physical situation. In a typical case, for a
quantum computer of $S$ qubits which interacts with the environment in
a thermal equilibrium, the off--diagonal elements of the density
matrix decay exponentially fast at a rate $\gamma S$~\cite{PES95}
\begin{equation}
 \rho_{ij}(t) \sim \rho_{ij}(0) e^{-\gamma St}
\end{equation}
where $\gamma=1/\tau_{dec}$ is a constant that describes the coupling
of a single qubit with the environment: the stronger the coupling, the
higher $\gamma$ and the smaller the decoherence time $\tau_{dec}$.

\begin{figure}
\vspace{4mm}
\centerline{
\psfig{width=5.56in,file=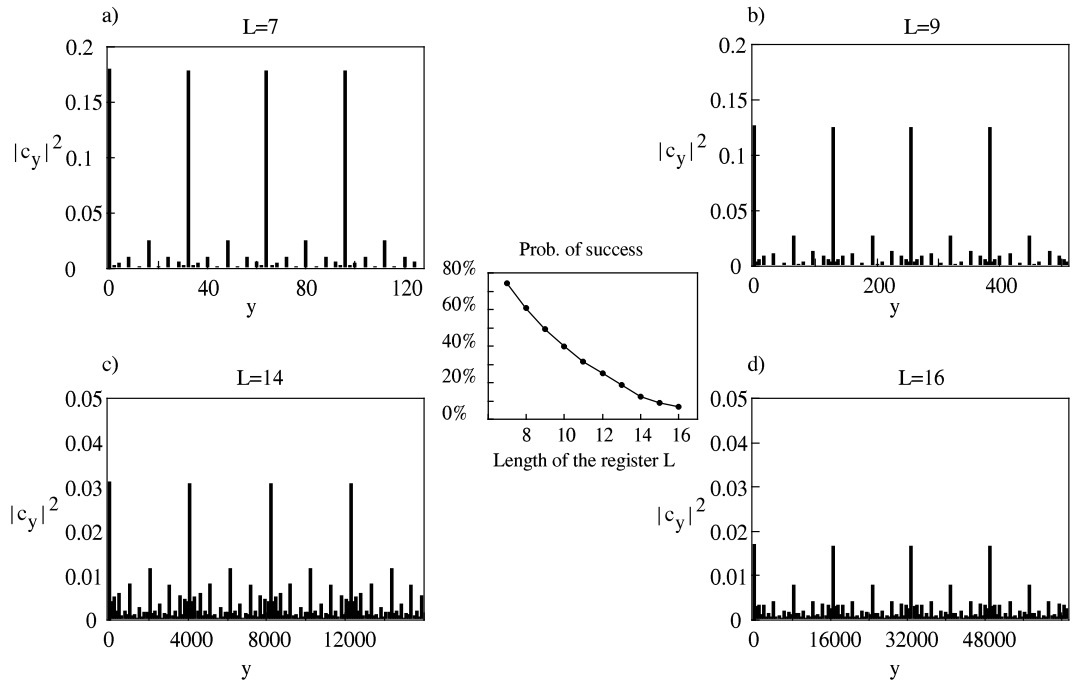}
}
\vspace{2mm}
\caption[fo7]
{\small Numerical simulations that mimic the effect of decoherence on
  the result of a {\small \sf DFT}. The initial state (not shown) is a
  periodic state with $r=4$ on a register of varying size (see
  Fig.~\ref{cxcy} for the case $L=8$). Without decoherence, the
  resulting state should have only $4$ components (Fig.~\ref{cxcy}b).
  The coupling with the environment induces errors (a-d), and reduces
  the probability of measuring the right answer. For a fixed $\gamma$
  and increasing $L$, the probability of getting the right peak
  decreases in a characteristic exponential way (central plot). The
  four plots a), b), c) and d) show the diagonal elements of the
  density matrix of the output. The intensity of the four principal
  peaks decreases as $L$ increases, and the probability of measuring a
  correct result decreases in an exponential fashion (central plot).
  Note that the scales on the vertical axis are adjusted for graphs c)
  and d).  }
\label{decoherenceDFT}
\end{figure}

For an efficient computation, we have seen that both $S$ and the total
computation time $t_{tot}$ required to complete the algorithm should
not grow faster than a polynomial, so that one can write
\begin{equation}
  S \sim L^\alpha \qquad t_{tot} \sim L^\beta t_{elem},
\end{equation}
where $t_{elem}$ is the characteristic time needed to perform a single
elementary computational step of the algorithm. From this, it is then
possible to show that the probability $\cal P$ of measuring the right answer at
the end of the quantum computation decreases exponentially with $S$
and $t_{tot}$, and hence with $L^{\alpha+\beta}$:
\begin{equation}
 {\cal P} \simeq e^{-\gamma S t_{tot}}=e^{-\gamma t_{elem} L^{\alpha+\beta}}.
\end{equation}
This can be illustrated in a very simple situation. Consider
performing a \DFT on a register of $L$ qubits that encodes a
superposition of period $r=4$. Without decoherence, we expect,
according to Eq.~(\ref{afterdft}), to measure with equal probability
either $\ket{0}$, $\ket{2^{L-2}}$, $\ket{2\cdot2^{L-2}}$ or $\ket{3
  \cdot 2^{L-2}}$. The measurement outcome will be affected by
decoherence and Fig.~\ref{decoherenceDFT} illustrates how the diagonal
elements of the density matrix of the state ({\em i.e.\/} the
probability outcomes of a measurement) behave. The calculation is
repeated for different $L$ with the same amount of decoherence (given
by a fixed $\gamma$).

To obtain at least one successful computation, one needs to run the
computer on average $1/{\cal P}=e^{\gamma t_{elem} L^{\alpha+\beta}}$
times.  Thus the problem becomes exponentially difficult as soon as
some decoherence is present. From the complexity point of view, the
magnitude of $\gamma$ has no relevance: as soon as there is some
coupling with the environment and $\gamma$ is non--zero, any
computation becomes inefficient. It is however quite clear that for
small $\gamma$ (long decoherence time $\tau_{dec}=1/\gamma$), it is
possible to effect some quantum operations before decoherence takes
its toll.  Technological progress in isolating quantum computers from
the environment and reducing decoherence will increase the largest
number that can be factored by such computers. The requirement for a
coherent computation to be completed within the decoherence time can be
written as
\begin{equation}
t_{tot}<\tau_{dec}/S=\tau_{dec}/L^\alpha.
\end{equation}
The right--hand side is the characteristic decoherence time of $L^\alpha$
qubits. From the above it follows that
\begin{equation}
  L<\left(\frac{\tau_{dec}}{t_{elem}}\right)^{1/(\alpha+\beta)}.
\end{equation}
With the best implementation of the factorisation algorithm $\alpha=1$
and $\beta=3$~\cite{VBE96}, hence the size of the largest number that
can be factored is bounded by $L<(\tau_{dec}/t_{elem})^{1/4}$. This is
an optimistic estimate in which only decoherence is taken into
account.  A careful analysis shows that this bound is dramatically
reduced when decay phenomena (such as spontaneous emission) are also
included~\cite{PK96}.

The ratio ${\cal M} = \tau_{dec}/t_{elem}$ is a useful figure of merit
for comparing different technologies. It tells us, very approximatively,
how many elementary operations can in principle be performed on a
single qubit before it decoheres.  Table~\ref{FOM} summarises
these values for some of the  technologies that could be used to implement
basic quantum computations. Some of these have already been used to
implement fundamental two--bit gates (such as the \CNOT
operation)~\cite{exper}. It is however still too early to say if the
present experimental setups can be scaled up easily and if these
technologies can be used to implement computations on many qubits.

\begin{table}
\centerline{
\begin{tabular}{l|c|c|c}
\hline 
Technology& $t_{elem}$ & $\tau_{dec}$  & $\cal M$  \\
\hline \hline 
M\"ossbauer nucleus & $10^{-19}$& $10^{-10}$ &$10^9$  \\
Electrons GaAs$^\star$ & $10^{-13}$& $10^{-10}$ &$10^3$ \\
Electrons Au & $10^{-14}$& $10^{-8}$ &$10^6$  \\
Trapped ions$^\star$ & $10^{-14}$& $10^{-1}$ &$10^{13}$  \\
Optical cavities & $10^{-14}$& $10^{-5}$ & $10^{9}$     \\
Electron spin & $10^{-7}$& $10^{-3}$ &$10^4$  \\
Electron quantum dot$^\star$ & $10^{-6}$& $10^{-3}$ &$10^3$  \\
Nuclear spin & $10^{-3}$& $10^{4}$ &$10^7$  \\
Superconductor islands$^\star$ & $10^{-9}$ & $10^{-3}$ & $10^6$  \\
\end{tabular}
}
\caption[t2]
{\small Figure of merit $\cal M$ for different technologies. The table
  gives respectively the characteristic decoherence time $\tau_{dec}$,
  the (minimum) time to complete an elementary operation $t_{elem}$
  (defined as $t_{elem}\simeq \hbar/\Delta E$, where $\Delta E$ is the
  energy splitting between the two states $\ket{0}$ and $\ket{1}$ of
  the qubit) and the number of operations $\cal M$ that could in
  principle be effected on a qubit before decoherence takes over.
  Asterisks indicate that the numbers given are to a large extent
  speculative.  The values are taken from~\cite{DV95}.}
\label{FOM}
\end{table}

\section{Conclusion}

As was the case in quantum cryptography a few years ago, the field of
quantum computation is rapidly growing and has already begun to move
from a theoretical to an experimental phase. Whether a full quantum
computer will ever be built remains an open question. The
technological challenge is immense and many problems must be
overcome first. A quantum computation requires  coherent quantum
evolution on a macroscopic scale. Not only that, it also needs to
be actively controlled. At present, we know of very few macroscopic
quantum states that involve many particles, or subsystems ({\em
  e.g.\/} electrons in a superconductor, $^4$He atoms in a superfluid,
or the recently observed Bose--Einstein condensates of Rb, Na or Li
atoms); in each case, the state is globally quantum and the particles
are not controlled on an individual basis. We may argue that each of
these quantum systems performs a quantum computation in its own, but
it may not be a computation we are interested in and moreover, we have
no real control over it.

The fragility of these macroscopic quantum systems tells us much about
the effect of decoherence. 
As explained in the last section, decoherence ultimately cuts out any
``exponential'' speed--up that we may gain from using quantum
computers and quantum algorithms, simply because to overcome it, the
best technique so far is to run the same computation over and over
again an exponential number of times (until we get a correct answer).
Fortunately, this is not the end of the story.  Classical computers
suffer from similar problems, and yet, one tends to agree that
classical computers are (in general!)  reliable.  This is because in
the classical situation we have efficient ways to fight errors.  For
existing computers, {\em error--correcting\/} codes have been designed
that are exponentially effective and that can handle and control
possible errors.  Unfortunately, these techniques cannot be directly
translated to tame decoherence in quantum computers. They are based on
redundancy ({\em i.e.\/} several  bits encoding one bit of information),
and more importantly, on a periodic monitoring of the state of the
computer.  Periodic monitoring means, {\em grosso modo\/}, measuring
the state of the computer, diagnosing an eventual error (and
eventually correcting it); but as we have seen, the mere act of
performing a measurement on a quantum computer will alter its state.
This obliges us to rethink the problem of error--correction in quantum
terms. First promising steps in this direction have already been
made~\cite{steane}.

From a fundamental point of view, it is irrelevant whether or not a
``quantum personal computer'' will materialise in the next decade.
More important is the insight we will gain in their study. Quantum
computation already tells us a lot  about the deep connections between
physics, computation and more generally information theory. No doubt
there is much more to learn.

\subsubsection*{Acknowledgements}

I would like to thank P.L.~Knight, M.~Abouzeid, T.~Brun, A.~Ekert,
C.~Macchiavello, A.~Sanpera, and K.--A.~Suominen for useful and critical
comments during the preparation of this manuscript. I am supported by
the Berrow fund at Lincoln College, Oxford, which I acknowledge
gratefully.

\appendix

\section{Discrete Fourier Transform}
\label{appendDFT}

Let us consider the simplified situation where the period $r$ divides
$2^L$ exactly.  A register in a periodic state is given for instance by
 Eq.~(\ref{firstreg}), 
which we rewrite
\begin{equation}
  \ket{\phi_{in}}=\sqrt{\frac{r}{2^L}}\;  
\sum_{j=0}^{G}  \ket{jr+l}
=  \sum_{x=0}^{2^L-1} c_x \ket{x}
\label{simple}
\end{equation} 
with $G=2^L/r$.  Performing a \DFT on $\ket{\phi_{in}}$ gives
\begin{equation}
  \ket{\phi_{out}} = \sum_y c_y\ket{y},
\end{equation}
where the amplitude of $c_y$ is
\begin{equation}
  c_y=\frac{\sqrt r}{2^L}\sum_{j=0}^{G}\exp\left(\frac{2\pi i
    (jr+l)y}{2^L}\right)= \frac{\sqrt
    r}{2^L}\exp\left(2\pi i\frac{l y}{2^L}\right)
 \left[ \sum_{j=0}^{G} \exp\left(2\pi i\frac{jry}{2^L}\right)\right].
\end{equation}
The term in the square bracket on the r.h.s. is zero unless $y$ is a
multiple of $2^L/r$,
\begin{equation}
  c_y = \left\{ 
\begin{array}{ll} 
\exp (2\pi i ly/2^L)/\sqrt{r} & \mbox{if $y$ is a multiple of $2^L/r$:
  $y=k2^L/r$ }\\ 
0 & \mbox{otherwise}
\end{array} 
\right.
\label{amp}
\end{equation}
Therefore, in the particular case when $r$ divides  $2^L$ exactly,
$\ket{\phi_{out}}$ can be written
\begin{equation}
  \ket{\phi_{out}} = \frac{1}{\sqrt{r}} \sum_{k=0}^{r-1} \exp (2\pi i
  lk/r) \; \ket{k 2^L/r}.
\label{afterdft2}
\end{equation}

\begin{figure}
\vspace{4mm}
\centerline{
\psfig{width=5.56in,file=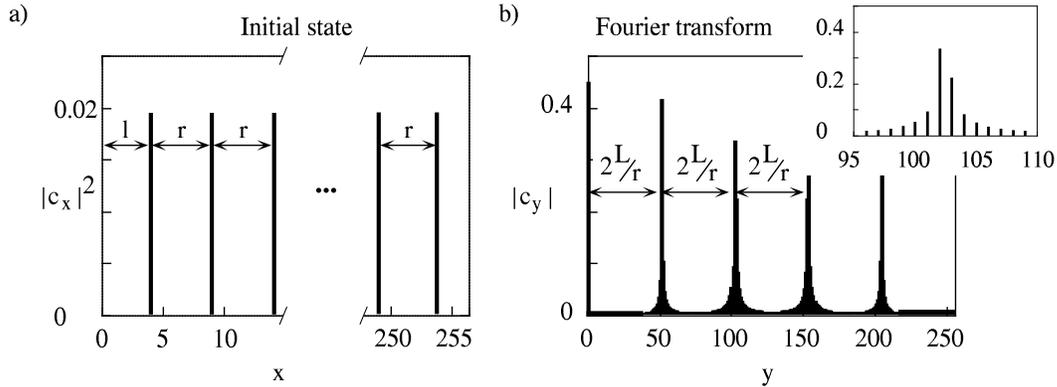}
}
\vspace{2mm}
\caption[fo8]
{\small Same as Fig.~\ref{cxcy}, but in this case $r=5$ does not
  divide exactly $2^L=256$. This results in a broadening of the peaks
  of the output state $\sum_y c_y \ket{y}$. Nevertheless, it can be
  shown that, when effecting a measurement on this state, the closest
  integers from multiples of $2^L/r$ are the most likely outcomes. This
  is illustrated in the inset: $102$ (local maximum), is the closest
  integer of $2 \times 2^8/5=102.4$. (Normally one plots $|c_y|^2$,
  {\em i.e.\/}, the actual probabilities, but here $|c(y)|$ is plotted
  to emphasise the spread of the peaks.)}
\label{nonexactcxcy}
\end{figure}

A more elaborate analysis is actually required when $r$ is not a
multiple of $2^L$. In this case, after effecting the {\small \sf DFT},
the coefficients $c_y$ are peaked on the closest integers to the
multiples of $2^L/r$ ({\em cf.\/} Fig.~\ref{nonexactcxcy}). These
peaks have a spread that decreases exponentially with $L$ (hence the
reason to choose in the factorisation algorithm the size of the first
register to be 2L).  A careful  analysis of this case can
be found in~\cite{EJ96}.

\newpage 

\begin{thebibliography}{99}

\bibitem{fundamentals} R.~Landauer, IBM J.~Res.~Dev. {\bf 5}, 183
  (1961); C.H.~Bennett, IBM J. Res. Dev. {\bf 17}, 525 (1973);
  C.H.~Bennett, Int.~J.~Theor.~Phys. {\bf 21}, 905 (1982);
  C.H.~Bennett, SIAM J.~Comput. {\bf 18(4)}, 766 (1989).

\bibitem {D85} D. Deutsch, D.,  Proc.~R.~Soc. London~A {\bf
    400}, 97 (1985).

\bibitem{S94} P.W. Shor, in {\it Proceedings of the 35th Annual
    Symposium on the Foundations of Computer Science}, edited by
  S.~Goldwasser (IEEE Computer Society Press, Los Alamitos, CA), p.
  124 (1994).

\bibitem{RSA} R. Rivest, A. Shamir, L. Adleman, {\it On Digital
    Signatures and Public--Key Cryptosystems\/}, MIT Laboratory for
  Computer Science, Technical Report, MIT/LCS/TR-212 (January 1979).

\bibitem{complexity} D.~Welsh, {\it Codes and cryptography\/}, Clarendon
  Press, Oxford, (1988). H.~S.~Wilf, {\it Algorithms and
    complexity\/}, Prentice-Hall, Englewood Cliffs / Prentice-Hall
  International, London, (1986).

\bibitem{lenstra} A.K.~Lenstra, H.W.~Lenstra Jr., M.S.~Manasse, and
  J.M.~Pollard,  in {\it Proc. 22nd ACM Symposium on the Theory
    of Computing\/}, p.564 (1990).

\bibitem{knuth}  D.E. Knuth, {\it The Art of Computer Programming, Volume 2:
    Seminumerical Algorithms} (Addison-Wesley, 1981).

\bibitem{S95} B. Schumacher, Phys.~Rev.~A {\bf 51}, 2738 (1995).


\bibitem{qdetect} A.S. Holevo, Problemy Peredachi Informatsii, {\bf
    9}, 3 (1979) (this journal is translated by IEEE under the title
  {\it Problems of Information Transfer\/}); E.B. Davies, IEEE Trans.
  Inform. Theory, {\bf IT 24}, 596 (1978); C.A. Fuchs and C.M. Caves,
  Phys.~Rev.~Lett. {\bf 73}, 3047 (1994).

\bibitem{qcrypt} S. Wiesner, Sigact News {\bf 15(1)}, 78 (1983);
  C.H.~Bennett and G.~Brassard in {\sl Proceedings of the IEEE
    International Conference on Computers, Systems, and Signal
    Processing, Bangalore, India\/} pp175, (IEEE, New York, 1984);
  A.~Ekert, Phys.~Rev.~Lett. {\bf 71}, 4287 (1993). For a review of
  the field, see also R.J.~Hughes, D.M.~Alde, P.~Dyer, G.G.~Luther,
  G.L.~Morgan and M.~Schauer, Contemporary Physics {\bf 36}(3), 149
  (1995); and also S.J.D.~Phoenix and P.D.~Townsend, Contemporary
  Physics {\bf 36}(3), 165 (1995).

\bibitem{BBCJPW93} C.H.~Bennett, G.~Brassard, C.~Cr\'epeau, R.~Jozsa,
  A.~Peres, and W.K.~Wootters, Phys.~Rev.~Lett. {\bf 70}, 1895 (1993).

\bibitem{BW92} C.H.~Bennett and S.~Wiesner, Phys. Rev. Lett. {\bf 69},
  2881 (1992).

\bibitem{peres} A. Peres, {\it Quantum Theory: Concepts and Methods}
  (Kluwer, 1993).

\bibitem{hardy} G.H. Hardy and E.M. Wright: {\sl An Introduction to the
    Theory of Numbers} (4th edition, Oxford University Press, 1965).

\bibitem{M76} G.L. Miller, Journal of Computer Science {\bf 13}, 300
(1976); 

\bibitem{EJ96} A.~Ekert and R.~Jozsa, Rev. Mod. Phys. {\em to appear 1996\/}.


\bibitem{D89} D. Deutsch, Proc.~R.~Soc. London~A {\bf 425}, 73
  (1989).

\bibitem{margolus} N.~Margolus, in {\it Complexity, Entropy, and the
    Physics of Information}, edited by W.~Zurek (Addison-Wesley) 1990;
  M.~Biafore, MIT Ph.D. Thesis (1993).
 

\bibitem{twobit} A.~Barenco, Proc. R. Soc. London A {\bf 449}, 679
  (1995); D.P. DiVincenzo, Phys. Rev. A {\bf 50}, 1015 (1995);
  T.~Sleator and H.~Weinfurter, Phys. Rev. Lett. {\bf 74}, 4087
  (1995).

\bibitem{universal} D.~Deutsch, A.~Barenco and A.~Ekert, Proc. R. Soc.
  London A {\bf 449}, 669 (1995); S.~Lloyd, Phys. Rev. Lett. {\bf 75},
  346 (1995).

\bibitem{gang9SD95} A.~Barenco, C.H.~Bennett, R.~Cleve, D.P.~DiVincenzo,
  N.~Margolus, P.~Shor, T.~Sleator, J.~Smolin and H.~Weinfurter,
  Phys.~Rev.~A {\bf 52}, 3457 (1995); J.A.~Smolin and D.P.~DiVincenzo,
  {\em Five Two-Bit Quantum Gates are Sufficient to Implement the
    Quantum Fredkin Gate\/}, preprint 1995.

\bibitem{VBE96} V. Vedral, A. Barenco and A. Ekert, {\em Quantum Networks
  for Elementary Operations\/}, submitted to Phys.~Rev.~A.

\bibitem{C94}  D. Coppersmith, IBM Research Report No. RC19642 (1994).
  D.~Deutsch, unpublished.

\bibitem{BETS96} A. Barenco, A. Ekert, P. T\"orm\"a and K.--A.
  Suominen, {\em On quantum implementation of the discrete Fourier
    transform and related algorithms\/}, submitted to Phys.~Rev.~A.

\bibitem{BN94} M. Brune, P. Nussenzveig, F. Schmidt-Kaler, F.  Bernardot,
  A. Maali, J.M. Raimond and S. Haroche, Phys.~Rev.~Lett. {\bf 72},
  3339 (1994).

\bibitem{BDEJ95} A.~Barenco, D.~Deutsch, A.~Ekert and R.~Jozsa,
  Phys.~Rev.~Lett. {\bf 74}, 4083 (1995).


\bibitem{ions} J.I. Cirac and P. Zoller, Phys.~Rev.~Lett {\bf 74},
  4091 (1995).

\bibitem{caltech} Q.A.~Turchette, C.J.~Hood, W.~Lange, H.~Mabuchi and
  H.J.~Kimble, {\em Measurement of conditional phsae shifts for
    quantum logic\/}, Caltech preprint.

\bibitem{cohen} See for example C.~Cohen-Tannoudji, B.~Diu,
  F.~Lalo\"e, {\em Quantum Mechanics} (Hermann and John Wiley \& Sons,
  1977).

\bibitem{devoret} M.~Devoret, private communication.

\bibitem{zurek} See for instance W.H. Zurek, Physics Today, {\bf
    44}(10), 36 (1991).
 
\bibitem{PES95} W.G.~Unruh, Phys.~Rev.~A {\bf 51}, 992 (1995); G.M.
  Palma, K.--A. Suominen and A.~Ekert, 1995, {\em Decoherence in
    quantum registers\/}, to be published in Proc.~R.~Soc.~London A.
  This phenomenon has been known for a while in other contexts, see
  for instance S.M. Barnett and P.L.~Knight, Phys.~Rev.~A {\bf 33},
  2444 (1986) for an example in quantum optics.

\bibitem{PK96} M. Plenio and P.L. Knight, {\em Realistic Lower Bounds
    for the Factorisation Time of Large Numbers on a Quantum
    Computer\/}, submitted to Phys.~Rev.~A.

\bibitem{exper}  C.~Monroe {\em et al.\/}, Phys.~Rev.~Lett.  {\bf 75},
  4714 (1995).

\bibitem{DV95} D. DiVincenzo, Phys. Rev. A, {\bf 50}, 1015 (1995).

\bibitem{steane} A.~Steane, {\em Multiple Particle Interference and
    Quantum Error Correction\/}, submitted to Proc. R. Soc. London A;
  A.R.~Calderbank and P.W.~Shor, {\em preprint.\/}; see also
  P.W.~Shor, Phys.~Rev.~A {\bf 52}, R2493 (1995).

\end{thebibliography}
\end{document}